\documentclass[aps,prl,twocolumn,superscriptaddress]{revtex4-2}
\usepackage{xspace}
\usepackage{graphicx}
\usepackage{xcolor}
\usepackage{hyperref}
\usepackage{booktabs}
\usepackage{mathptmx}
\usepackage{dcolumn}     
\usepackage{bm}
\usepackage{upgreek}
\usepackage{tabularx}

\usepackage[mathlines,running]{lineno}

\newcommand{\nnbb}    {\ensuremath{2\nu\beta\beta}\xspace}
\newcommand{\onbb}    {\ensuremath{0\nu\beta\beta}\xspace}
\newcommand{\nuc}   [2]{$^{#2}$#1\xspace} 
\newcommand{\Ge}      {$^{76}$Ge\xspace}
\newcommand{\kgyr}    {{kg\,yr}\xspace}

\newcommand{\bbHL}    {\ensuremath{\smash{T_{1/2}^{2\nu}}}\xspace}

\newcommand{\effnmes} {\ensuremath{{\mathcal{M}^{2\nu}_{\text{eff}}}}\xspace}
\newcommand{\phsp}    {\ensuremath{G^{2\nu}}\xspace}
\newcommand{\qbb}     {\ensuremath{Q_{\beta\beta}}\xspace}
\newcommand{\gerda}   {\textsc{Gerda}\xspace}
\newcommand{\geant}   {\textsc{Geant4}\xspace}
\newcommand{\mage}    {\textsc{MaGe}\xspace}
\newcommand{\PI}      {Phase~I\xspace}
\newcommand{\PII}     {Phase~II\xspace}

\begin{document}

\title{Final Results of \gerda on the Two-Neutrino Double-$\beta$ Decay Half-Life of \Ge}

\collaboration{{\textsc{Gerda} collaboration}}
\email{correspondence: gerda-eb@mpi-hd.mpg.de}
\noaffiliation
  \affiliation{INFN Laboratori Nazionali del Gran Sasso, Assergi, Italy}
  \affiliation{INFN Laboratori Nazionali del Gran Sasso and Gran Sasso Science Institute, Assergi, Italy}
  \affiliation{INFN Laboratori Nazionali del Gran Sasso and Universit{\`a} degli Studi dell'Aquila, L'Aquila,  Italy}
  \affiliation{INFN Laboratori Nazionali del Sud, Catania, Italy}
  \affiliation{Institute of Physics, Jagiellonian University, Cracow, Poland}
  \affiliation{Institut f{\"u}r Kern- und Teilchenphysik, Technische Universit{\"a}t Dresden, Dresden, Germany}
  \affiliation{Joint Institute for Nuclear Research, Dubna, Russia}
  \affiliation{European Commission, JRC-Geel, Geel, Belgium}
  \affiliation{Max-Planck-Institut f{\"u}r Kernphysik, Heidelberg, Germany}
  \affiliation{Department of Physics and Astronomy, University College London, London, UK}
  \affiliation{INFN Milano Bicocca, Milan, Italy}
  \affiliation{Dipartimento di Fisica, Universit{\`a} degli Studi di Milano and INFN Milano, Milan, Italy}
  \affiliation{Institute for Nuclear Research of the Russian Academy of Sciences, Moscow, Russia}
  \affiliation{Institute for Theoretical and Experimental Physics, NRC ``Kurchatov Institute'', Moscow, Russia}
  \affiliation{National Research Centre ``Kurchatov Institute'', Moscow, Russia}
  \affiliation{Max-Planck-Institut f{\"ur} Physik, Munich, Germany}
  \affiliation{Physik Department, Technische  Universit{\"a}t M{\"u}nchen, Germany}
  \affiliation{Dipartimento di Fisica e Astronomia, Universit{\`a} degli Studi di 
Padova, Padua, Italy}
  \affiliation{INFN  Padova, Padua, Italy}
  \affiliation{Physikalisches Institut, Eberhard Karls Universit{\"a}t T{\"u}bingen, T{\"u}bingen, Germany}
  \affiliation{Physik-Institut, Universit{\"a}t Z{\"u}rich, Z{u}rich, Switzerland}
%
%
\author{M.~Agostini}
  \affiliation{Department of Physics and Astronomy, University College London, London, UK}
\author{A.~Alexander}
  \affiliation{Department of Physics and Astronomy, University College London, London, UK}
\author{G.R.~Araujo}
  \affiliation{Physik-Institut, Universit{\"a}t Z{\"u}rich, Z{u}rich, Switzerland}
\author{A.M.~Bakalyarov}
  \affiliation{National Research Centre ``Kurchatov Institute'', Moscow, Russia}
\author{M.~Balata}
  \affiliation{INFN Laboratori Nazionali del Gran Sasso, Assergi, Italy}
\author{I.~Barabanov}
  \affiliation{Institute for Nuclear Research of the Russian Academy of Sciences, Moscow, Russia}
\author{L.~Baudis}
  \affiliation{Physik-Institut, Universit{\"a}t Z{\"u}rich, Z{u}rich, Switzerland}
\author{C.~Bauer}
  \affiliation{Max-Planck-Institut f{\"u}r Kernphysik, Heidelberg, Germany}
\author{S.~Belogurov}
  \altaffiliation[also at: ]{NRNU MEPhI, Moscow, Russia}
  \affiliation{Institute for Theoretical and Experimental Physics, NRC ``Kurchatov Institute'', Moscow, Russia}
  \affiliation{Institute for Nuclear Research of the Russian Academy of Sciences, Moscow, Russia}
\author{A.~Bettini}
  \affiliation{Dipartimento di Fisica e Astronomia, Universit{\`a} degli Studi di Padova, Padua, Italy}
  \affiliation{INFN  Padova, Padua, Italy}
% ...
\collaboration{\textsc{Gerda} collaboration}
\noaffiliation


\begin{abstract}
  We present the measurement of the two-neutrino double-$\beta$ decay rate of \Ge performed with the \gerda \PII experiment. 
  With a subset of the entire \gerda exposure, 11.8\,\kgyr, the half-life of the process has been determined: 
  $\bbHL = (2.022 \pm 0.018_{\text{stat}} \pm 0.038_{\text{sys}}) \times 10^{21}$\,yr.
  This is the most precise determination of the \Ge two-neutrino double-$\beta$ decay half-life and one of the most precise measurements 
  of a double-$\beta$ decay process. The relevant nuclear matrix element can be extracted: 
   $\mathcal{M}^{2\nu}_{\text{eff}} = (0.101\pm 0.001). $
\end{abstract}


\maketitle

The two-neutrino double-$\beta$ (\nnbb) decay is a rare nuclear transition in which two neutrons are simultaneously transformed into two protons, and two electrons and two antineutrinos are created, ensuring lepton number conservation. It is among the rarest radioactive processes ever detected. It has been observed in several nuclei with half-lives ranging between 10$^{18}$--10$^{24}$\,yr~\cite{Barabash:2020nck}. Double-$\beta$ decay transitions are a unique probe for particle physics.  The discovery of neutrino-less double-$\beta$ (\onbb) decay, in which no neutrinos are emitted, would reveal the Majorana nature of neutrinos and have a strong connection with the nature of the neutrino mass generation mechanism~\cite{Schechter:1981bd}. Observing \onbb decay would provide evidence for lepton number violation and directly point to physics beyond the Standard Model~\cite{DellOro:2016tmg}.
Several extensions of the Standard Model also predict the emission of exotic particles as a byproduct instead of two antineutrinos. A popular hypothetical decay mode involves Majoron emission~\cite{Hirsch:1995in}, but various other candidate particles have been proposed~\cite{Agostini:2020cpz, Bolton:2020ncv}. Experimental searches for new physics in double-$\beta$ decay transitions rely on sophisticated nuclear matrix element calculations to convert decay rates to information on the underlying particle physics model~\cite{Engel:2016xgb}.  In this context, measured \nnbb decay rates can be used as a test bench to validate and improve nuclear structure calculations~\cite{Barea:2015kwa, Rodin:2007fz, Ferreira:2017bxx, Jokiniemi:2022ayc}.

\begin{figure}[tb]
    \centering
    \includegraphics[width=\columnwidth]{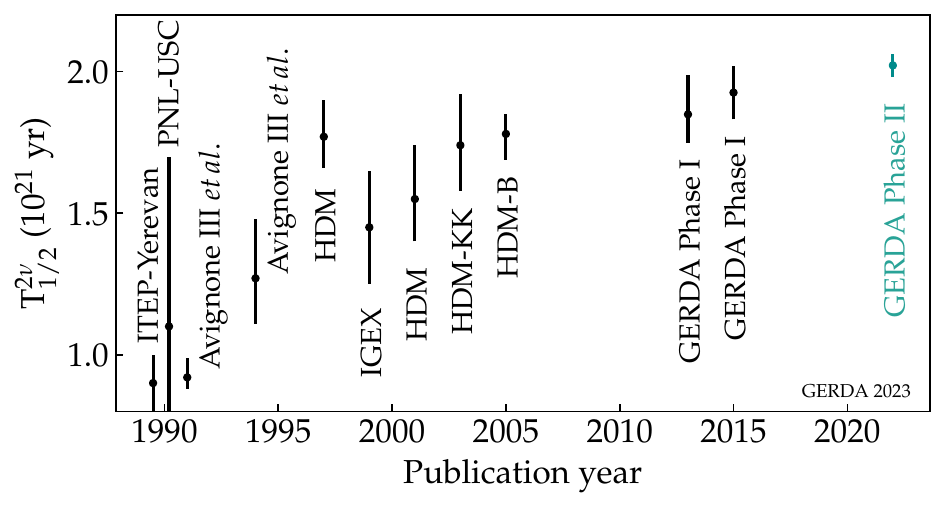}
    \caption{%
      History of published \Ge \nnbb decay half-life measurements~\cite{Vasenko:1989jv, Miley:1990er, Avignone:1991bk, Avignone:1994kn, Gunther:1997ai, KlapdorKleingrothaus:2000sn, Morales:1998hu, Dorr:2003gf, Bakalyarov:2003jk, Agostini:2012nm, Agostini:2015nwa}.
  }
    \label{fig:half_life_ge}
\end{figure}

In this Letter, we present the final measurement of the \nnbb decay half-life (\bbHL) of \Ge performed with the \gerda experiment. 
Pioneering measurements of this quantity were performed already in the nineties, while the most recent result was reported by \gerda \PI~\cite{Agostini:2015nwa}, 
which measured a half-life of $\bbHL = (1.926 \pm 0.095) \times 10^{21}$~yr. A collection of measurements performed over the years is reported in Fig.~\ref{fig:half_life_ge}.
An increase of the \bbHL central value is observed. It has been attributed to a systematic underestimation of the background, which decreases as the experiments keep reducing their background level~\cite{Agostini:2012nm}. 
The precision of previous \gerda \PI measurements was limited by systematic uncertainties related to the fit model and the detector's active mass. Both sources of uncertainties have been drastically reduced in \gerda \PII through a re-determination of the active volume for a selection of germanium detectors and the utmost reduction of the background by detecting the scintillation light produced by background events depositing energy in liquid argon (LAr), leading to the results described in this Letter.

The \gerda experiment was located underground at the Laboratori Nazionali del Gran Sasso (LNGS) of INFN, in Italy~\cite{GERDA:2012qwd, Agostini:2017hit, Agostini:2019hzm}. High-purity germanium detectors built from material isotopically enriched in \Ge were operated inside a 64\,m$^3$ LAr cryostat~\cite{Knopfle:2022fso}. In the second phase of the experiment, 7 coaxial and 30 Broad Energy Germanium (BEGe) detectors were mounted in 6 strings~\cite{Agostini:2017hit}, each enclosed in a transparent nylon vessel to prevent the collection of radioactive potassium ions on the detector surfaces~\cite{Lubashevskiy:2017lmf}. A curtain of wavelength-shifting fibers connected to silicon photomultipliers and low-activity photomultiplier tubes were arranged around the detector array. This instrumentation allowed for effective detection of the argon scintillation light due to background events depositing energy in the argon surrounding the germanium detectors~\cite{Agostini:2017hit}.  The LAr cryostat was submerged in a 590\,m$^3$ water tank instrumented with photomultipliers and used, together with scintillator panels on the top of the setup, for the muon veto system~\cite{Freund:2016fhz}.

The experimental signature of a \nnbb decay is a well-localized energy deposition within a germanium detector.  The total decay energy is shared among the two electrons and two antineutrinos produced in the process. The electrons release all their energy in germanium within a few millimeters from the decay location. The antineutrinos escape the detector carrying away a fraction of the energy. Thus, the detectable energy varies between zero and the Q-value of the reaction, $\qbb = 2039.061(7)$\,keV~\cite{Mount:2010xyz}, with a maximum around 700\,keV. Several background sources can also generate events in this energy range~\cite{gerda-lar-model}.
Up to about 565\,keV, the event rate of \gerda is dominated by the $\beta$ decay of \nuc{Ar}{39}, a cosmogenic isotope of argon. Above this threshold, the majority of the detected events is due to \nnbb decays with minor contributions from \nuc{Ac}{228}, \nuc{Th}{228}, \nuc{Bi}{214}, \nuc{Co}{60}, and \nuc{K}{40} in structural material; \nuc{K}{42} decays in the LAr surrounding the detectors; $\alpha$ decays on the p$^+$ electrode of the detectors.

The analysis presented in this work is based on data collected during the \PII of the project, processed following the procedures and digital signal processing algorithms described in~\cite{Agostini:2019hzm}.  Unphysical events due to electrical discharges or noise and data collected in periods of hardware instabilities are identified and removed from the data set using the methods discussed in~\cite{Agostini:2020xta}. Events accompanied by a light signal in the water tank or LAr are also discarded.
The energy deposited within the detectors is reconstructed using a zero-cusp-area filter~\cite{Agostini:2015pta} which provides a resolution better than 3\,keV full-width half maximum with BEGe detectors over a wide energy range extending up to \qbb~\cite{Agostini:2021duc}.

\begin{figure}
    \centering
    \includegraphics[width=\columnwidth]{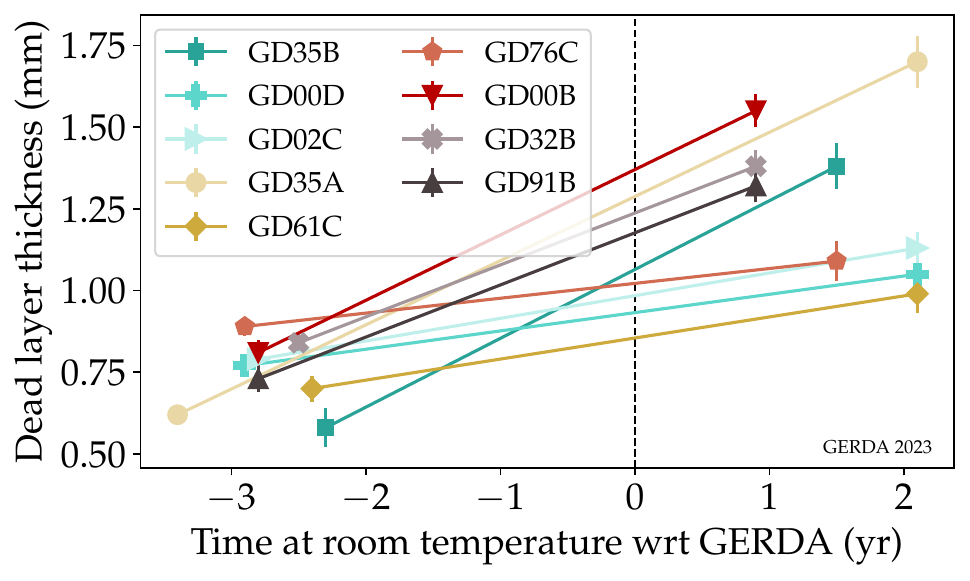}
    \caption{Dead layer thickness (DLT) measured 
    before the beginning and after the end of the \gerda data taking shown as a function of the time the detectors were stored at room temperature. The whole \gerda data taking is collapsed at zero since no growth is expected while the detectors are operated in liquid argon.}
    \label{fig:dead_layer_growth_and_interpolation}
\end{figure}

Of the 103.7\,\kgyr of exposure collected in \gerda \PII~\cite{Agostini:2020xta}, only data collected with nine BEGe detectors between December 2015 and April 2018 have been used in this work, corresponding to a total exposure of 11.8\,\kgyr.
These detectors were chosen because they have been characterized before their deployment in the \gerda LAr cryostat~\cite{Agostini2019} and after the end of the \gerda data taking, allowing for the determination of the dead layer thickness (DLT), and, in turn, of the active volume fraction ($f_{AV}$), during the \gerda data taking. 
The DLT is defined as the distance from the detector surface at which charge carriers are fully collected and is known to grow at room temperature. No reliable model of this process has been formulated yet. Different growths were observed among the nine re-characterized BEGe detectors, as shown in Fig.~\ref{fig:dead_layer_growth_and_interpolation}. The two measurements were performed with the same apparatus to reduce systematic uncertainties~\cite{Agostini2019}.
A linear interpolation between the two measured values is used as an estimation of the DLT during \gerda data taking. The interpolated DLT values are summarized in Table~\ref{tab:9BEGe_dataset_summary}. The uncertainties reported here are obtained by propagating the uncertainties of the two measured values to the linear interpolated value.
The obtained DLT was used to determine the $f_{AV}$, {\it i.e.} the fraction of the entire detector volume where an energy deposition is fully reconstructed.
The resulting $f_{AV}$ values are also summarized in Table~\ref{tab:9BEGe_dataset_summary}.
\begin{table}
    \centering
    \caption{Summary of the nine BEGe detectors used in this work. The individual analysis exposures, the dead layer thickness (DLT) values, and the corresponding active volume fractions ($f_{AV}$) are reported. Uncertainties have been calculated assuming a linear growth of the DLT with time. 
    }
    \label{tab:9BEGe_dataset_summary}
    \vspace*{10pt}
    \begin{tabular*}{0.8\columnwidth}{@{\extracolsep{\fill} } cccc }
        \toprule
        Detector & Exposure  & DLT& $f_{AV}$ \\ 
        name & (kg\,yr) & (mm) & \\ \midrule
        GD35B & 1.6 & 1.02 $\pm$ 0.10 & 0.888 $\pm$ 0.010\\
        GD00D & 1.5 & 0.86 $\pm$ 0.08 & 0.904 $\pm$ 0.009\\
        GD02C & 1.5 & 0.93 $\pm$ 0.08 & 0.897 $\pm$ 0.009\\
        GD35A & 1.5 & 1.25 $\pm$ 0.09 & 0.868 $\pm$ 0.009\\
        GD61C & 1.1 & 0.80 $\pm$ 0.09 & 0.900 $\pm$ 0.010\\
        GD76C & 1.6 & 0.96 $\pm$ 0.09  & 0.895 $\pm$ 0.010\\
        GD00B & 1.3 & 1.29 $\pm$ 0.11 & 0.850 $\pm$ 0.013\\
        GD32B & 1.4 & 1.13 $\pm$ 0.10 & 0.872 $\pm$ 0.011\\
        GD91B & 0.5 & 1.10 $\pm$ 0.11 & 0.871 $\pm$ 0.013 \\ \bottomrule
    \end{tabular*}
\end{table}
While it was possible to determine the active volume of the nine re-characterized BEGe detectors with an accuracy of 1--1.5\%, the active volume of the other BEGe detectors and that of the coaxial detectors is not known at the same degree of accuracy. For this reason, data from the latter detectors are excluded from the analysis. The active volume ultimately determines the detection efficiency for \nnbb decays. Therefore, its uncertainty directly translates into a systematic uncertainty on the \nnbb decay half-life. In contrast, the statistical uncertainty expected using only 11.8\,\kgyr of exposure is subdominant. 
Data collected after the upgrade of the experimental setup in the summer of 2018  with the same nine BEGe detectors were excluded from the analysis due to major changes in the LAr instrumentation that are not included in the modeling of the LAr veto system~\cite{gerda-lar-model}. While the loss of exposure is minimal, the LAr veto system and its Monte Carlo simulation are crucial elements of the analysis,
as will be explained in the following.

The statistical analysis follows the methods used in our previous work, where limits on different exotic double-$\beta$ decays have been set~\cite{gerda-exotics}.
A binned maximum likelihood fit is used to estimate the number of \nnbb and background events. The fit is performed in the energy window between 560 and 2000\,keV, using a 10\,keV binning. It was checked that the fit results were not affected by the binning. 
The free parameters of the fit are the normalization factors of the signal and background distributions. Thus, the normalization factor of the \nnbb decay distribution corresponds to the number of \nnbb decay events observed in the data set, $N_{2\nu}$. All the parameters of the fit are unconstrained. The statistical inference relies on a frequentist approach and the profile-likelihood ratio test statistic~\cite{Zyla:2020zbs}. The only parameter of interest is $N_{2\nu}$, while all normalization factors of the background distributions are treated as nuisance parameters, and their uncertainties are propagated by profiling. The test statistic distributions are evaluated with Monte Carlo techniques, generating a set of \gerda pseudo-experiments assuming different signal hypotheses. These distributions are used to extract the 68\% probability interval on the number of \nnbb counts. 

\begin{figure}
    \includegraphics[width=\columnwidth]{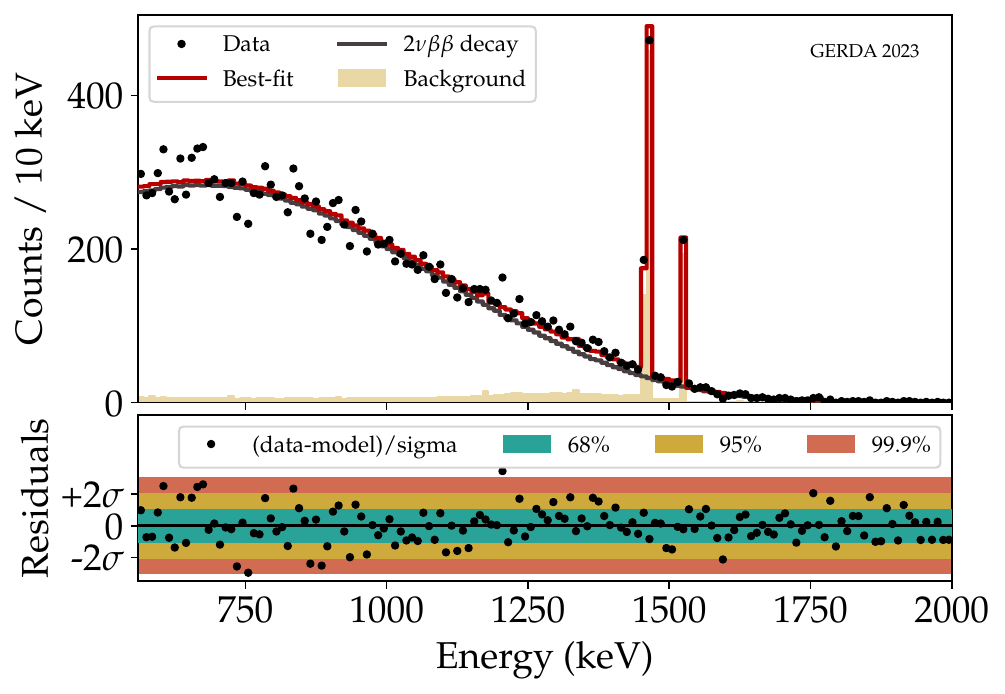}
    \caption{%
      Best-fit background and signal decomposition of the energy distribution
      of 11.8\,\kgyr of data from \gerda \PII after applying the liquid
      argon veto cut. In the bottom panel, the difference between data and model normalized over the expected Poisson standard deviations is shown together with 68\%, 95\%
      and 99\% confidence intervals.
    }\label{fig:fit_results}
\end{figure}

In this analysis' scope, reducing the background to a minimal level is crucial, and the LAr veto cut plays a fundamental role in achieving this. A background model after the LAr veto cut has been developed in~\cite{gerda-lar-model} and used in this work, as in our previous one~\cite{gerda-exotics}. It includes separate model components for \nuc{Ac}{228}, \nuc{Th}{228}, \nuc{Bi}{214}, \nuc{Co}{60}, and \nuc{K}{40} decays, two model components for \nuc{K}{42} decay, one for the decay in the LAr volume and one for the decay very close to detector surfaces, and finally, a linear distribution for the decay of $\alpha$ particles on the p$^+$ electrode surface. 
To obtain the probability distributions of signal and background sources after the LAr veto cut, the simulation of the LAr scintillation light production and detection chain was implemented in the \geant-based simulation framework \mage~\cite{Boswell:2010mr}. The determination of the LAr veto condition for Monte Carlo events is described in detail in~\cite{gerda-lar-model}. 

Fig.~\ref{fig:fit_results} shows the experimental data and the total best-fit model. The contributions to the total fit model of the \nnbb decay and the background are also shown separately. The residual background after the LAr veto cut is extremely low, and the signal-to-background ratio, excluding the two prominent $\gamma$-lines from \nuc{K}{40} and \nuc{K}{42}, is 22:1, while it was only 2:1 in the same energy range according to the background model before analysis cut~\cite{Abramov:2019hhh}. Thus, the LAr veto cut reduces the background of more than a factor 10 in the energy
region dominated by the \nnbb decay, as already pointed out in~\cite{gerda-lar-model,gerda-exotics}. The residuals are shown at the bottom of Fig.~\ref{fig:fit_results} in units of standard deviation. Their distribution is compatible with a Gaussian distribution centered at 0 and with a width equal to 1. 

The best-fit value and 68\% probability interval on the number of \nnbb\ counts, extracted from the observed test statistic, 
is $N_{2\nu} = (16911 \pm 147_{\text{stat}} \pm 112_{\text{sys}})$ counts in the fit range. 
The systematic uncertainty here accounts for the contributions that affect the energy distribution of signal and background and, 
in turn, the estimation of $N_{2\nu}$ from the fit. These are folded into the analysis during the computation of the test statistic distribution 
(See prior-predictive method in Refs.~\cite{Zyla:2020zbs,Demortier2007}). Each time a pseudo-experiment is generated, a new generative model is 
created by sampling the model parameters from prior distributions. As a result of this procedure, the tail of the test statistic is widened, and the 
systematic uncertainties are naturally incorporated in the result of the statistical inference.

The energy distribution of the background depends on the location of the background contamination. Different peak-to-Compton values are expected for $\gamma$ decays very close to the detector or far from them, as well as for events depositing energy in the bulk volume or close to the surface. As stated above, the background model after the LAr veto cut includes a minimal set of locations for the background components~\cite{gerda-lar-model}. In the Monte Carlo generation of the pseudo-experiments, the location of each background contribution is uniformly sampled among all the locations identified in~\cite{Abramov:2019hhh}. The resulting systematic uncertainty on the determination of $N_{2\nu}$ is $\pm0.62\%$, as summarized in Table~\ref{tab:sys_unc}. 

\begin{table}
  \caption{%
    Summary of the systematic uncertainties affecting the \nnbb decay half-life
    estimate.
  }\label{tab:sys_unc}
  \vspace*{10pt}
  \begin{tabular}{r@{\hspace{2em}}l}
    \toprule
    Source                 & Uncertainty \\
    \midrule
    Background model & $\pm 0.62\%$ \\
    Liquid argon veto model          & $\pm 0.21\%$ \\
    n$^+$ detector contact model     & $< 0.1\%$   \\
    Theoretical \nnbb decay model    & $\pm 0.13\%$   \\
    [5pt]
    \emph{Sub Total (fit model)}     & $\pm 0.66\%$ \\
    [5pt]
    Active volume                    & $\pm 1.8\%$ \\
    Enrichment fraction              & $\pm 0.3\%$ \\
    [5pt]
    \emph{Total}                     & $\pm 1.9\%$ \\
    \bottomrule
  \end{tabular}
\end{table}

The response of the LAr veto instrumentation also affects the shape of the background probability distributions. Uncertainties in the optical parameters used for the Monte Carlo simulation affect the probability of detecting the scintillation light depending on the point where the emission takes place in the LAr. The uncertain parameters include, among others, the LAr attenuation length and the reflectivity of materials in the detector array. Macroscopic properties of the background distributions, such as the peak-to-Compton ratio, are modified by these parameters. The systematic uncertainty in the LAr veto response is heuristically parametrized with methodologies discussed in~\cite{gerda-lar-model}. The reader is referred to the latter publication for a complete treatment of the topic. The observed systematic uncertainty due to LAr veto model uncertainties on the $N_{2\nu}$ value is $\pm0.21\%$, as summarized in Table~\ref{tab:sys_unc}. It is worth remarking that, because of the very localized topology of \nnbb decay events, the corresponding event distribution depends neither on the geometry nor on the LAr veto response.

A systematic uncertainty contribution is also expected from the modeling of germanium detectors. Every detector is characterized by a transition region between the active volume and the dead layer in which the charge-collection efficiency is partial~\cite{Lehnert2016}. The efficiency profile and the size of this transition layer affect the shape of signal and background probability distributions, particularly the low energy region of the \nnbb decay and the lower tail of intense $\gamma$ peaks. A linear efficiency profile for the transition layer, whose average size is about 50\% of the full dead layer region for BEGe detectors~\cite{Lehnert2016}, is used as default. Still, different profiles are considered in the systematic uncertainties, and the size varied in a conservative range of $\pm5$ standard deviations from the central value. Nevertheless, the overall effect on $N_{2\nu}$ is less than 0.1\%, smaller than the other contributions summarized in Table~\ref{tab:sys_unc}.

The theoretical calculations for the shape of the \nnbb decay of \Ge assume Higher-State Dominance (HSD), i.e.~the hypothesis that all intermediate states of the intermediate nucleus contribute to the decay rate~\cite{Kotila:2012zza}. Assuming the alternative Single-State-Dominance (SSD) hypothesis, i.e.~the \nnbb decay is governed by a virtual two-step transition through the first 1$^+$ state of the intermediate nucleus~\cite{Domin:2004za}, a tiny difference in the shape of the \Ge \nnbb decay is observed~\footnote{Private communication with J.~Kotila and F.~Iachello.}. This difference is maximal in the tail of the \nnbb decay distribution where the statistic is very low but less than 0.5\% at the peak of the distribution. This results in a $\pm$0.13\% systematic uncertainty on the determination of $N_{2\nu}$. 


$N_{2\nu}$ is converted into the decay half-life (\bbHL) through the relation
\[
  \bbHL = \frac{1}{N_{2\nu}}\cdot \frac{\mathcal{N}_A\, \ln(2)}{m_{76}} \, \varepsilon \, \mathcal{E} \;,
\]
where $\mathcal{N}_A$ is the Avogadro's constant, $m_{76}$ the molar mass of the enriched germanium, $\mathcal{E}$ the exposure, and $\varepsilon$ the total efficiency of detecting \nnbb decays in the analysis range. The latter includes the electron containment efficiency, the active volume fraction, the \Ge enrichment fraction, and the efficiency of the analysis cuts. 

To determine the systematic uncertainty on the \bbHL due to the uncertainty of the detector active volume fractions, we sum the nine volumes using the DLT values measured before data taking (upper boundary) and the DLT values measured after data taking (lower boundary). We randomly sample the total active volume uniformly between the lower and upper boundaries and take the RMS of the resulting distribution as a systematic uncertainty. Hence, the uncertainties of the DLT values are treated completely correlated among the nine detectors and do not rely on any assumption on the time profile of the DLT growth rate during storage at room temperature. This conservative estimate yields a relative uncertainty of 1.8\%, larger than the single $f_i^{AV}$ uncertainties given in table~\ref{tab:9BEGe_dataset_summary}.

The \Ge enrichment fraction was estimated to be ($87.4\pm0.3$)\% for the BEGe detectors and contributes with a 0.3\% relative uncertainty on \bbHL. 
This uncertainty is smaller than reported previously due to a re-evaluation of the estimates documented in~\cite{Agostini2019}.
All the contributions to the systematic uncertainty budget are summarized in Table~\ref{tab:sys_unc}. Uncertainties about other efficiency factors, such as the containment efficiency and the efficiency of the analysis cuts, are negligible. 

From the number of \nnbb decay events evaluated from the fit and the systematic uncertainties listed above, we obtain
\[
  \bbHL = (2.022 \pm 0.018_{\text{stat}} \pm 0.038_{\text{sys}}) \times 10^{21}\,\text{yr}\;.
\]
Summing in quadrature statistical and systematic uncertainties, the total $1\sigma$ uncertainty on \bbHL is 2.1\%. 
This is dominated by the systematic uncertainty on the active volume (1.8\%). The total contribution to the systematic uncertainty from the fit model is only 0.7\%, 
comparable to the 0.9\% statistical uncertainty. 

\begin{figure}
    \centering
    \includegraphics[width=\columnwidth]{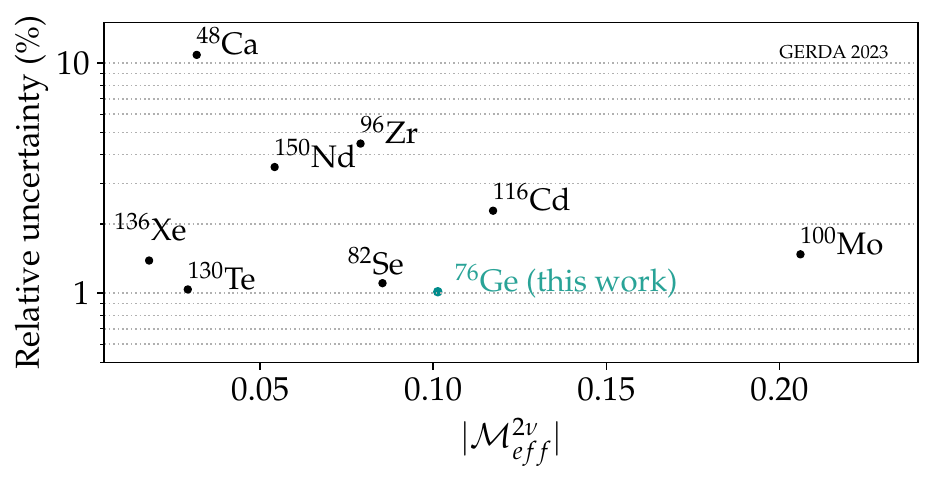}
    \caption{Experimental values of the effective nuclear matrix elements $|{\effnmes}|$ for various double-$\beta$ emitters with the precision given on the vertical axis. 
    The different $|{\effnmes}|$ were calculated using for each isotope the most precise determination of the 
    half-life to date~\cite{Adams2021, Albert:2013gpz, Azzolini:2019yib, Armengaud:2019rll, Barabash:2018yjq, Arnold:2016ezh, Arnold:2016qyg, Argyriades:2009ph} 
    and the phase space factors from~\cite{Kotila:2012zza}.}
    \label{fig:comparison_diff_isotopes}
\end{figure}

\bbHL is converted into an experimental estimation of the effective nuclear matrix elements \effnmes through the relation
\[
  [\bbHL]^{-1} = \phsp\, |{\effnmes}|^2 \;,
\]
where \phsp\ is the phase space factor. This has been calculated for \Ge with sub-percent accuracy $\phsp = 48.17\times 10^{-21}\,\text{yr}^{-1}$~\cite{Kotila:2012zza}.  
With the \bbHL\ extracted in this work, the effective nuclear matrix element is ${\effnmes}=(0.101\pm0.001)$. 
A comparison of the experimental values of \effnmes obtained with different isotopes is shown in Fig.~\ref{fig:comparison_diff_isotopes}. 
Despite the longer half-life of \Ge compared to other isotopes (up to two orders of magnitude), the result obtained in this work for \Ge aligns with the high precision 
reached in the last years by several experiments and represents one of the most precise measurements of a double-$\beta$ decay process. 
Present calculations of these nuclear matrix elements yield a precision that is far off that achieved in experiments.

In conclusion, we performed the most precise determination of the \Ge \nnbb decay half-life: $\bbHL = (2.022 \pm 0.042) \times 10^{21}\,\text{yr}\;$.
The half-life measured in this work is compatible with the past \gerda \PI results. It confirms the trend of slightly increasing \bbHL\ central value over time as the experiments progressively increase the signal-to-background ratio. The superior signal-to-background ratio achieved in this work is the result of the extremely low background condition in which \gerda operated combined with the excellent background rejection capabilities of the LAr veto system. The unprecedented precision in the determination of the \Ge \nnbb decay half-life benefits from that and from the precision determination of the active volume of the BEGe detectors. In fact, the statistical uncertainty is subdominant even using only a limited exposure, while the systematic uncertainty related to the active volume of the germanium detectors dominates the total uncertainty. Further improvement of the precision of the \Ge \nnbb decay half-life estimate would require a precision determination of the active volume of the germanium detectors, which will be the challenge of the LEGEND experiment, the future of double-$\beta$ decay physics with \Ge~\cite{LEGEND:2021bnm}. 

The data shown in Figs.~\ref{fig:dead_layer_growth_and_interpolation} and~\ref{fig:fit_results} are available in
ASCII format as Supplemental Material~\cite{supp-mat}.

\begin{acknowledgments}
We would like to thank F.~Iachello and J.~Kotila for the useful discussions and for providing the calculations of the \nnbb energy spectrum under the different theoretical assumptions considered in this work. 

The \textsc{Gerda} experiment is supported financially by
the German Federal Ministry for Education and Research (BMBF),
the German Research Foundation (DFG),
the Italian Istituto Nazionale di Fisica Nucleare (INFN),
the Max Planck Society (MPG),
the Polish National Science Centre (NCN, grant number UMO-2020/37/B/ST2/03905), 
the Polish Ministry of Science and Higher Education (MNiSW, grant number DIR/WK/2018/08),
the Russian Foundation for Basic Research,
and the Swiss National Science Foundation (SNF).
This project has received funding/support from the European Union's
\textsc{Horizon 2020} research and innovation programme under
the Marie Sklodowska-Curie grant agreements No 690575 and No 674896.
This work was supported by the Science and Technology Facilities Council, part of the U.K. Research and Innovation (Grant No. ST/T004169/1).
The institutions acknowledge also internal financial support.
 
The \textsc{Gerda} collaboration thanks the directors and the staff of the LNGS for their continuous strong support of the \gerda experiment.

\end{acknowledgments}


%

\end{document}